\documentclass{ws-procs9x6}

\newcommand\beq{\begin{eqnarray}}
\newcommand\eeq{\end{eqnarray}}
\newcommand\be{\begin{eqnarray}}
\newcommand\ee{\end{eqnarray}}
\newcommand{\bea}{\begin{eqnarray}}
\newcommand{\eea}{\end{eqnarray}}

\newcommand\kT{{\bf k}_T}
\newcommand{\kt}{{\bf k}_T^2}

\def\xbj{x_{\mbox{\tiny B}}}
\begin{document}

\title{Spin-Orbit Correlations and Single-Spin Asymmetries}

\author{M. BURKARDT}

\address{Dept. of Physics, New Mexico State University,
Las Cruces, NM 88003, USA\\
E-mail: burkardt@nmsu.edu\thanks{This work was supported by the DOE
({\sc DE-FG03-95ER40965})}}


\maketitle

\abstracts{Several examples for the role of orbital angular 
momentum and spin-orbit correlations in hadron structure are 
discussed.}
\section{Introduction}
In nonrelativistic quantum mechanics, Fourier
transforms of the form factors yield charge
distributions in the center of mass frame.
In general, the concept of a center of mass 
has no analog in relativistic theories, and 
thus the position space interpretation of form
factors is frame dependent. This is different in
the Infinite Momentum Frame (IMF) or light-cone framework, where
a Galilean subgroup of transverse boosts allows introducing
the transverse center of longitudinal momentum as the weighted
average of transverse (i.e. $\perp$ to the boost direction)
positions of all partons, weighted by their momentum fractions. This center of $\perp$ momentum is the
reference point in the interpretation of the (two dimensional)
Fourier transform of Generalized Parton Distributions (GPDs) as 
Impact Parameter dependent parton Distributions (IPDs) 
[\refcite{mb1}]. A similar interpretation exists for the 2-d 
Fourier transform of form factors [\refcite{soper}] as the
latter can also be obtained by integrating GPDs
over the momentum fraction.

The distribution of partons in impact parameter also plays a role
for Single-Spin Asymmetries (SSAs): as one expects the final state 
interactions on the ejected quark in a Semi-Inclusive
Deep-Inelastic Scattering (SIDIS) experiment
to be on average attractive, any sideward deformation of IPDs
is expected to result in an enhancement of the transverse
momentum distribution of the ejected quark in the opposite
direction. This observation forms the basis for a qualitative
link between GPDs and SSAs [\refcite{mb:SSA}]. 

\section{Impact Parameter Dependent PDFs and the Sivers Effect}
\label{sec:IPDs}
The Fourier transform of the GPD 
$H_q(x,0,t)$ yields the  
distribution $q(x,{\bf b}_\perp)$ of 
unpolarized quarks, for an unpolarized target, in 
impact parameter space  
\be
q(x,{\bf b}_\perp)= 
 \int \!\!\frac{d^2{\bf \Delta}_\perp}{(2\pi)^2}  
H_q(x,0,\!-{\bf \Delta}_\perp^2) \,e^{-i{\bf b_\perp} \cdot
{\bf \Delta}_\perp}, \label{eq:GPD}
\ee 
with ${\bf \Delta}_\perp = {\bf p}_\perp^\prime -{\bf p}_\perp$.
For a transversely polarized target (e.g. when polarized in the
$+\hat{x}$-direction) the impact parameter dependent
PDF (IPD) $q_{+\!\hat{x}}(x,{\bf b}_\perp)$ is
no longer axially symmetric and
the transverse deformation is described
by the gradient of the Fourier transform of the GPD $E_q(x,0,t)$
\bea
q_{+\!\hat{x}}(x,\!{\bf b_\perp}) 
&=& q(x,\!{\bf b_\perp})
- 
\frac{1}{2M} \frac{\partial}{\partial {b_y}} \!\int \!
\frac{d^2{\bf \Delta}_\perp}{(2\pi)^2}
E_q(x,0,\!-{\bf \Delta}_\perp^2)\,
e^{-i{\bf b}_\perp\cdot{\bf \Delta}_\perp}
\label{eq:deform}
\eea
$E_q(x,0,t)$ and hence the details of this deformation are not 
very well known, but its $x$-integral, the Pauli form factor
$F_2$, is. This allows to
relate the average transverse deformation resulting from Eq.
(\ref{eq:deform}) to the contribution from the
corresponding quark flavor to the anomalous magnetic moment.
This observation is important in understanding the sign of the
Sivers function.

In a target that is polarized transversely ({\it e.g.} vertically), 
the quarks in the target 
nucleon can exhibit a (left/right) asymmetry of the distribution 
$f_{q/p^\uparrow}(\xbj,{\bm k}_T)$ in their transverse 
momentum ${\bm k}_T$ [\refcite{sivers,trento}]
\be
f_{q/p^\uparrow}(\xbj,{\bm k}_T) = f_1^q(\xbj,k_T^2)
-f_{1T}^{\perp q}(\xbj,k_T^2) \frac{ ({\bm {\hat P}}
\times {\bm k}_T)\cdot {\bm S}}{M},
\label{eq:sivers}
\ee
where ${\bm S}$ is the spin of the target nucleon and
${\bm {\hat P}}$ is a unit vector opposite to the direction of the
virtual photon momentum. The fact that such a term
may be present in (\ref{eq:sivers}) is known as the Sivers effect
and the function $f_{1T}^{\perp q}(\xbj,k_T^2)$
is known as the Sivers function.
The latter vanishes in a naive parton 
picture since $({\bm {\hat P}} \times {\bm k}_T)\cdot {\bm S}$ 
is odd under naive time reversal (a property known as naive-T-odd), 
where one merely reverses
the direction of all momenta and spins without interchanging the
initial and final states. The momentum fraction $x$, 
which is equal to $\xbj$ in DIS experiments, 
represents the longitudinal momentum of the quark {\it before}
it absorbs the virtual photon, as it is determined solely from the 
kinematic properties of the virtual photon and the target nucleon. 
In contradistinction, the transverse momentum ${\bm k}_T$ is 
defined in terms of the kinematics
of the final state and hence it represents the asymptotic 
transverse momentum
of the active quark {\it after} it has left the target and before it
fragments into hadrons. Thus the Sivers function for semi-inclusive
DIS includes the final state interaction 
between struck quark and target remnant, and
time reversal invariance no longer requires that it vanishes.
\begin{figure}
\unitlength1.cm
\begin{picture}(10,2.5)(2.8,14)
\includegraphics{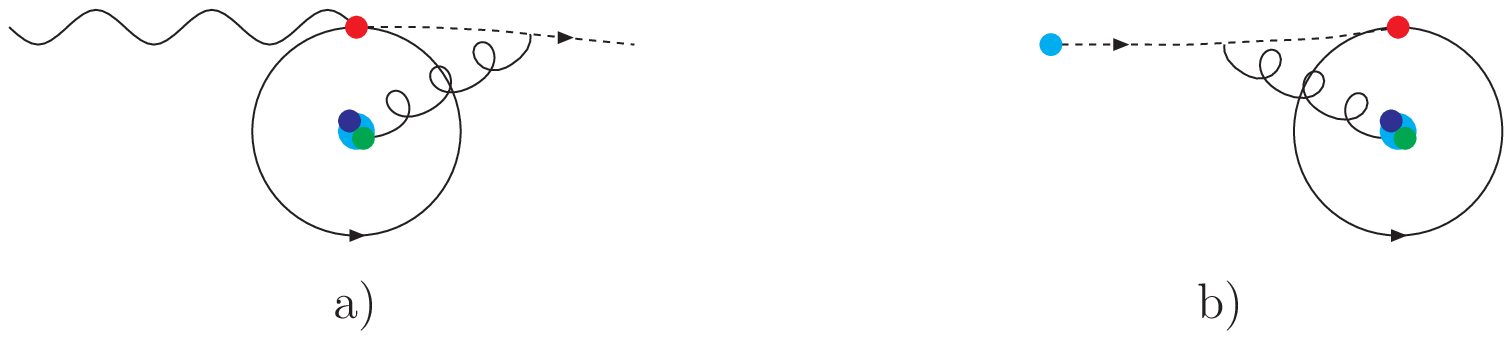}
\end{picture}
\caption{In SIDIS (a) the ejected (red) quark is attracted by
the (anti-red) spectators. In contradistinction, in DY (b), 
before annihilating with the (red) active quark, the approaching
(anti-red) antiquark is repelled by the (anti-red) spectators.}
\label{fig:SIDISDY}
\end{figure}
Indeed, as time reversal not only reverses the signs of all
spins and momenta, but also transforms final state interactions (FSI)
into initial state interactions (ISI), 
it has been shown that the Sivers 
function relevant for SIDIS and that relevant for 
Drell-Yan (DY) processes must have opposite signs 
[\refcite{collins}],
\be
f_{1T}^\perp(\xbj,{k}_T^2)_{SIDIS} =
- f_{1T}^\perp(\xbj,{k}_T^2)_{DY} ,
\label{SIDISDY}
\ee
where the asymmetry in DY arises from the ISI between the 
incoming antiquark and the target.
The experimental verification of this relation 
would provide a test of the current understanding of the Sivers 
effect within QCD. 
It is instructive to elucidate its physical
origin in the context of a perturbative picture:
for instance, when the virtual photon in a DIS process hits a red 
quark, the spectators must be collectively anti-red in order to
form a color-neutral bound state, and thus attract
the struck quark (Fig. \ref{fig:SIDISDY}). 
In DY, when an anti-red antiquark annihilates with
a target quark, the target quark must be red in order to merge
into a photon, which carries no color. Since the proton was 
colorless before the scattering, the spectators must be anti-red
and thus repel the approaching antiquark.

\begin{figure}
\unitlength1.cm
\begin{picture}(10,2.3)(3.,19.2)
\includegraphics{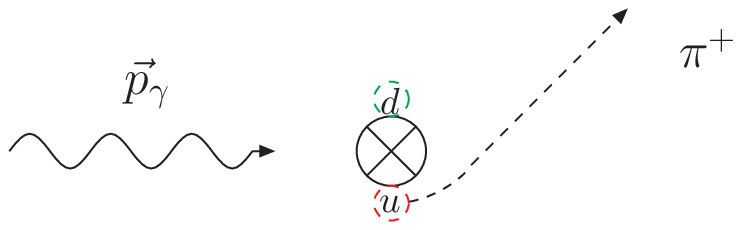}
\end{picture}
\caption{The transverse distortion of the parton cloud for a proton
that is polarized into the plane, in combination with attractive
FSI, gives rise to a Sivers effect for $u$ ($d$) quarks with a
$\perp$ momentum that is on the average up (down).}
\label{fig:deflect}
\end{figure}

The significant distortion of parton distributions in impact 
parameter space (\ref{eq:deform})
provides a natural mechanism for a Sivers effect.
In semi-inclusive DIS, when the 
virtual photon strikes a $u$ quark in a $\perp$ polarized proton,
the $u$ quark distribution is enhanced on the left side of the target
(for a proton with spin pointing up when viewed from the virtual 
photon perspective). Although in general the final state 
interaction (FSI) is very complicated, we expect it to be on average attractive thus translating a position space
distortion to the left into a momentum space asymmetry to the right
and vice versa (Fig. \ref{fig:deflect}).
Since this picture is very intuitive, a few words
of caution are in order. First of all, such a reasoning is strictly 
valid only in mean field models for the FSI as well as in simple
spectator models [\refcite{spectator}]. 
Furthermore, even in such mean field models
there is no one-to-one correspondence between quark distributions
in impact parameter space and unintegrated parton densities
(e.g. Sivers function). While both are connected by a Wigner
distribution [\refcite{wigner}], 
they are not Fourier transforms of each other.
Nevertheless, since the primordial momentum distribution of the quarks
(without FSI) must be symmetric we find a qualitative connection
between the primordial position space asymmetry and the
momentum space asymmetry (with FSI). 
Another issue concerns the $x$-dependence of the Sivers function.
The $x$-dependence of the position space asymmetry is described
by the GPD $E(x,0,-{\Delta}_\perp^2)$. Therefore, within the above
mechanism, the $x$ dependence of the Sivers function should be
related to the $x$ dependence of $E(x,0,-{\Delta}_\perp^2)$.
However, the $x$ dependence of $E$ is not known yet and we only
know the Pauli form factor $F_2=\int {\rm d}x E$. Nevertheless, 
if one makes
the additional assumption that $E$ does not fluctuate as a function 
of $x$ then the contribution from each quark flavor $q$ to the
anomalous magnetic moment $\kappa$ determines the sign of 
$E^q(x,0,0)$
and hence of the Sivers function. Making these assumptions,
as well as the very plausible assumption that the FSI is on average
attractive, 
one finds that $f_{1T}^{\perp u}<0$, while 
$f_{1T}^{\perp d}>0$. Both signs have been confirmed by a flavor
analysis based on pions produced in a SIDIS experiment 
by the {\sc Hermes} collaboration [\refcite{hermes}].

\section{Charge Density in the Center of the Neutron}

As integrating the GPD $H^q$ over the momentum fraction
$x$ of the active quark yields the Dirac form factor
$F_1^q$, integrating
Eq.(\ref{eq:GPD}) over  $x$ also
provides an interpretation of the Dirac form factor $F_1^q$ as
the 2d Fourier transform of the charge density
(from quarks with flavor $q$) in impact parameter space 
\be
\rho({\bf b}_\perp) = \int \frac{{\rm d}^2{\bf \Delta}_\perp}
{(2\pi)^2} e^{i{\bf b}_\perp \cdot {\bf \Delta}_\perp}
F_1^q(t=-{\bf \Delta}_\perp^2).
\label{eq:rho}
\ee
The main advantage of Eq. (\ref{eq:rho}) compared to the
Fourier transform of the Sachs form factors is that 
$\rho({\bf b}_\perp)$ has a density interpretation 
[\refcite{soper,mb:prob}].

Application of Eq.(\ref{eq:rho}) to $F_1$ for the
neutron [\refcite{millereffect}], 
yields a charge density $\rho({\bf b}_\perp)$ that
is negative not only at very large ${\bf b}_\perp$ but also
near ${\bf b}_\perp =0$. The negative charge
density at large distances ${\bf b}_\perp$ has the well known
interpretation in terms of the pion cloud through the virtual
process $n\rightarrow p\pi^-$, but the negative charge density
near the origin appears to be mysterious. 
The key for intuitively understanding the negative charge density
in the center of the neutron seems to be Orbital Angular Momentum
(OAM).

With the benefit of hindsight, the first evidence for the
presence of OAM in the nucleon wave function
came from the existence of a large anomalous magnetic moment
$\kappa_p=1.79$ and $\kappa_n=-1.91$.
Indeed, in a relativistic theory, an
anomalous magnetic moment necessarily implies the presence
of wave function components with nonzero OAM 
(for a detailed discussion, see e.g. Refs. 
[\refcite{gunar,gardner}]).
Further insight can be gained by performing a flavor
decomposition of the anomalous magnetic moment. Neglecting
the small contribution from strange and heavier quarks and making 
use of charge symmetry, one finds for the contribution from
$u$ and $d$ quarks
\be
\kappa^u_p =\kappa^d_n=1.67
\quad \quad \quad
 \kappa^d_p =\kappa^u_n=-1.91
\ee
respectively.
Here charge factors have been taken out such that for example
$\kappa_p=1.79 =\frac{2}{3}\kappa^u_p-\frac{1}{3}\kappa^d_p$. 
For the purpose of this paper, we observe that not only are
$\kappa^q_N$ large but that the magnitude of the
contribution from the 
minority flavor ($d$ in the proton and $u$ in the neutron) is even
slightly larger than that of the corresponding majority flavor.
Given that there are less down quarks in the proton, in combination
with the fact that a nonzero anomalous magnetic moment
requires wave function components with OAM,
this result suggests that a $d$ quark in a proton has a 
significantly higher
probability to be found with nonzero OAM than a $u$ quark.

The second piece of evidence comes from studies
of the Sivers function $f_{1T}^\perp$. A recent 
flavor analysis based on pions produced in SIDIS suggests
a nonzero Sivers function for both $u$ and $d$ quarks with
approximately equal magnitude and opposite sign 
$
f_{u/p^\uparrow} \approx -f_{d/p^\uparrow}
$ [\refcite{hermes}].
Again we note that even though the proton contains more $u$ than
$d$ quarks, the
Sivers function for $d$ quarks is comparable in magnitude with
those for $u$ quarks, again indicating that $d$ quark wave function
components have a larger $p$ wave component.

Finally, we turn our attention to recent lattice calculations.
Using the Ji relation 
$
J_q = \frac{1}{2} \int \!{\rm d}x \, x
\left[H_q(x,0,0)+E_q(x,0,0)\right]
$ 
to determine the contribution 
$J_q$ of quark flavor $q$ to the nucleon spin 
from the GPDs $H_q$ and $E_q$ [\refcite{JiPRL}],
and after subtracting the quark
spin contribution, one finds [\refcite{lattice}] 
\be
L_u \approx - L_d\approx 0.15 \label{eq:lattice}
\ee
i.e. about equal in magnitude and with opposite sign. While
it is not entirely clear how to relate the OAM obtained through 
the Ji relation, to the OAM in light-cone wave function
(the latter being relevant for the anomalous magnetic moment
and the Sivers function) this result confirms our observation
that the smaller number of $d$ quarks yields the same 
magnitude for the OAM, i.e. again a larger contribution
from each $d$ quark (in the proton).

Despite the fact that there are less $d$ than $u$ quarks in the 
proton, they contribute with about the same magnitude to
the anomalous magnetic moment, the Sivers function, and the
quark OAM from the Ji-relation.
These observations indicate that the wave function for $d$ quarks
(in a proton)
has a larger $p$-wave component than the one for
$u$ quarks. Charge symmetry implies that in 
a neutron $u$ quarks have a larger $p$ wave components than $d$
quarks. Since $p$ wave function components are suppressed at the 
origin, this naturally suppresses $u$ quarks in a neutron for small 
${\bf b}_\perp$ compared to an $SU(2)$ symmetric solution thus
providing a qualitative explanation for the surprising result 
from Ref. [\refcite{millereffect}]. 

\section{Tensor Correlations}
Another set of observables that are sensitive to spin-orbit 
correlations are Transverse Momentum dependent parton Distributions
(TMDs).
Projecting out quarks with transverse spin ${\bf s}$, the most general
expression for the $\kt$-dependence of parton distributions reads
[\refcite{AMS}]
\bea
q(x,\kT,{\bf s}, {\bf S}) &=& \frac{1}{2}\left[f_1 + s^iS^ih_1+
\frac{1}{M}S^i \varepsilon^{ij}k^j f_{1T}^\perp + 
\frac{1}{M}s^i \varepsilon^{ij}k^j h_{1}^\perp\right.\\
& &\left.\quad\quad
+ \frac{1}{M}\Lambda s^ik^i h_{1L}^\perp + \frac{1}{2M^2}s_iS_j
\left( 2k^ik^j - \kt\delta^{ij}\right) h_{1T}^\perp
\right], \nonumber
\eea
where $\Lambda$ is the longitudinal nucleon polarization 
and ${\bf S}$ its transverse spin.
Two more terms appear when one also considers longitudinally
polarized quarks.

In the following, we will focus on the 
chirally odd tensor correlation $h_{1T}^\perp$, which contributes to 
matrix elements with
a double spin asymmetry in orthogonal transverse directions.
In a helicity basis, this implies that it contributes to
matrix elements where both quark and nucleon helicities 
flip --- but in opposite directions --- resulting in
a helicity mismatch by two units. A specific example is
the transition from a nucleon state with 
$S_z=+\frac{1}{2}$ to $S_z=-\frac{1}{2}$ while the spin of the
active quark flips from $s_z=-\frac{1}{2}$ to $s_z=+\frac{1}{2}$.
The active quark thus has to absorb $L_z=-2$. 
The fact that $L_z$ changes by two units requires either the presence
of wave function components with $L_z=\pm 2$ ($s$-$d$ interference),
or matrix elements that are quadratic in the $p$ wave component.
In either case, applying the power-counting techniques from Ref.
[\refcite{feng}], this implies that 
$h_{1T}^\perp\stackrel{x\rightarrow 1}{\longrightarrow}(1-x)^5$.

We will focus on contributions
quadratic in the $p$ wave component and neglect $s$-$d$ 
interference. Consider a nucleon that is
polarized in the $+\hat{x}$ direction. When the active quark
has $l_x=\pm 1$, its distribution is enhanced in the 
$\hat{y}-\hat{z}$
plane. When viewed from the $\hat{z}$ direction, the
distribution is thus enhanced
along the $\hat{y}$ axis, but suppressed along the $\hat{x}$ axis
(imagine a bagel in the $\hat{y}-\hat{z}$ plane viewed from the
side). On the other hand, for $l_x=0$, the quark distribution is 
enhanced along the $\hat{x}$ axis (peanut aligned along with the
$\hat{x}$ axis). As
this deformation is described by $h_{1T}^\perp$, this TMD thus
appears the ideal tool to decide whether (and in which spin
configuration) the polarized quark density looks more like a bagel
or a peanut, or perhaps even a pretzel [\refcite{miller}].

In order to better understand the specific implications for the
nucleon's angular momentum structure, let us consider for
example the case $ h_{1T}^\perp<0$. According to the above
discussion, this case corresponds to $l_x=0$ when quark and nucleon
transversity are anti-parallel, and $l_x=\pm 1$ when they are 
parallel. In quark models the tensor correlation $h_{1T}^\perp$
usually arises solely from the lower component (standard representation for Dirac matrices) and for the lower
component transverse spin and transversity have opposite signs.
Thus the case $h_{1T}^\perp<0$ corresponds to 
$l_x=0$ when quark and nucleon
transverse spin are parallel, and $l_x=\pm 1$ when they are 
anti-parallel. In a bag model or potential model, this type
of correlation arises naturally when $j_q$ is in the same direction as
the nucleon spin, as the $p$ wave component arises from the lower 
component of the quark wave function, and is largest when
$j_q$ and $s_q$ are anti-parallel with $l_q$ parallel to $j_q$.
In most models, one would thus expect $h_{1T}^{\perp u}<0$.
The case $h_{1T}^\perp>0$
corresponds to $l_x=\pm 1$ when quark and nucleon
transversity are parallel (transverse spins anti-parallel), 
and $l_x=0$ when they are 
anti-parallel (transverse spins parallel). 
This naturally arises in quark states with
$l_q$ and $s_q$ coupled to a net $j_q$ that is oriented opposite to
the nucleon spin. We thus expect $h_{1T}^{\perp d}>0$ 
with smaller absolute magnitude than $h_{1T}^{\perp u}$.

In contradistinction to the Sivers and Boer-Mulders functions,
$h_{1T}^\perp$ is (naive) T-even and does not {\it require} the
presence of nontrivial ISI/FSI phases. Nevertheless, measurements
of $h_{1T}^\perp$ in SIDIS or DY experiments may still
be affected by ISI/FSI. While such effects may be small
compared to the intrinsic $h_{1T}^\perp$, it is not {\it a priori}
clear how to separate ISI/FSI from intrinsic effects.

In Ref. [\refcite{DH}] it has been shown that a tensor correlation
analogous to the one described by $h_{1T}^\perp$ in momentum space
is described by the Fourier transform of the GPD
$\tilde{H}_T^{''}$ in impact parameter space. 
There the issue of ISI/FSI does not arise, as
the GPDs are defined through matrix elements of local operators
and directly probe the intrinsic quark densities.
While it is not possible to directly map
a density in position space onto a density in momentum,
(single particle) $p$-orbits have the same angular distribution in 
momentum space
as they have in position space. Therefore, a bagel in momentum space
corresponds to a bagel in position space as well 
and one would thus expect that $h_{1T}^\perp$  
has the same sign as $\tilde{H}_T^{''}$.  
Generalizing the approach from Ref. \refcite{ZLu}, more concrete
relations between tensor correlation in momentum and impact-parameter
space have been derived in Ref. \refcite{Metz} for the specific
case of diquark models.

\end{document}